\documentclass{ws-p8-50x6-00}
\usepackage{overcite}

\begin{document}

\title{ON STRING GAS COSMOLOGY AT FINITE TEMPERATURE}

\author{MONICA BORUNDA}

\address{ISAS-SISSA, Via Beirut 2-4 and INFN,\\I-34013 Trieste, Italy\\E-mail: mborunda@sissa.it}

\maketitle

\abstracts{
  Within an adiabatic approximation, thermodynamical 
equilibrium and a small, nine dimensional, toroidal universe as
 initial conditions, 
we analyze the evolution of the dimensions in 
two different regimes: (i) the Hagedorn
regime, with a single scale factor with a nearly constant time evolution (ii) an almost-radiation 
dominated regime,  including the leading corrections due to the 
lightest Kaluza Klein and winding modes, in which for some initial conditions
the large dimensions continue to expand and the small ones
remain small.}

\section{Introduction}

In the last decades string theory  has become a promised candidate
for the underlying theory of the fundamental interactions of nature.
However, even though lots of progress has been done, it has not been
yet possible to confront it with real physics. One possibility to achieve this
is through cosmology by studying the cosmological implications of
string theory.
 On the other hand, string theory provides an alternative to answer the basic
questions that standard cosmology currently faces, such as the initial
singularity,  dimensionality of spacetime, cosmological constant,
horizon and flatness problem and the origin of the density perturbations
in the cosmic microwave background. All these reasons suggest that 
cosmology and strings complement each other in several ways, giving
rise to string cosmology.

A rigorous way to understand string cosmology is by studying the
time dependent backgrounds in string theory
\cite{{Horowitz:2002mw},{Craps:2002ii},{Liu:2002ft}},
 unfortunately
this is a hard task due to cosmological singularities
that arise in this context, 
  and still much work has to be done.
On the other hand, a good approximation can be done
 by studying the adiabatic evolution of
 time independent backgrounds in string theory.
In this direction lots of work have been done where
a low energy effective approach is made  (see  
\cite{{Quevedo:2002xw},{Lidsey:1999mc}}
 for recent reviews, discussions and
references),  
like for example, Brandenberger and Vafa (B\&V) 
model\cite{Brandenberger:1988aj} in 1988, continuing with the Pre-Big-Bang scenario\cite{Gasperini:2002bn},
 ekpyrotic universes\cite{Khoury:2001bz},   brane cosmology\cite{Binetruy:1999ut}, and so on.

Winding states and {\it T-duality}  are intrinsic features
  of string theory
 and  play a fundamental role in  the B\&V model, which main objectives
 are the  solution to the cosmological singularity problem and the
 dimensionality of spacetime.
 In this scenario, the universe starts as a 9 dimensional small
torus filled with an ideal gas of strings and evolves such that 6 dimensions 
remain small and 3 dimensions become large as nowadays. How does it take place?
Since we are dealing with a compact universe filled with a string gas, the
strings wind around the dimensions. When the system is out of thermal
equilibrium there is interaction of this states, in particular,
 the winding states wrapping around one sense can annihilate with the
states wrapping around the other sense, anti-winding states.
 Now, the key ingredient is that
a string describes a 2 dimensional world-sheet, therefore it is easy for the strings
to see each other when the dimensionality of space is 3 or less. On the other hand,
 when the dimensions are more than 3,  the strings generically miss each other being unable 
to annihilate.  This, in principle,  leads to 3 spatial dimensions free
to expand (due to the Kaluza Klein (KK) contribution), and on the other
hand, leave the remaining 6 spatial dimensions with winding strings
 around them preventing them from expansion. On the other hand, since
string theory exhibits {\it T-duality} ($R\rightarrow 1/R$),
it implies that neither the temperature nor the physical length are singular as
 $R\rightarrow 0$, avoiding the cosmological singularity.

Even though this work\cite{Bassett:2003ck}  is inspired by the previous
 dynamics, we will not
study them, but we will be interested in the adiabatic
 evolution  that
takes place when the universe is in thermal equilibrium filled with an
ideal gas of closed strings
 (so we are always dealing with weak string coupling). 
The analysis is performed in such a way that the ``string ingredients'',
{\it T-duality} and the presence of winding states, are manifestly present. 
 We study two regimes.
 First, we 
address the question of what happens if we start off with 9
 dimensional small toroidal
 universe within a micro-canonical string treatment, Hagedorn regime. 
It turns out that the dimensions remain nearly constant around the initial value.
Second, we  assume that by some mechanism, for example, B\&V,
there are $d$ large and $9-d$ small dimensions, almost radiation regime,
 and we  study their
evolution within a canonical string approach, finding out that for a set
of initial conditions it is possible to keep the small dimensions
small compared to the large ones (see also\cite{Watson:2003gf}
for a different approach leading to the same conclusion).

\section{Set up}
In this toy model we consider  type {\bf IIA}/{\bf IIB} closed strings,   on a 
9 dimensional torus. The starting point is  the low-energy effective action\cite{{Gasperini:1992em},{Veneziano:1991ek},{TV}},
 with a massless dilaton field which for simplicity is just considered time
dependent, $\phi =\phi (t) $, 
\begin{equation}
S=\int d^{10}x\sqrt{-g} [{\rm e}^{-2\phi}\{R+4(\nabla\phi )^2\} + {\cal L}_M], 
\end{equation}
where $g$ is the determinant of the background metric $g_{\mu \nu}$, and
${\cal L}_M$ corresponds to the matter contribution from the string gas. 
The metric is taken as $g_{\mu\nu}={\rm diag} (-1,R_1^2(t),\dots,R_9^2(t))$, 
with scale factors $R_i(t)={\rm e}^{\lambda_i(t)}$.
It then follows that the dilaton gravity equations of motion\cite{TV}
\begin{equation}
\begin{array}{rcl}
 -\sum_{i=1}^9\dot\lambda_i^2+\dot\psi^2&=&{{\rm e}^\psi} E \, ,\\
\ddot\lambda_i-\dot\psi\dot\lambda_i&=&\frac 12 {{\rm e}^\psi} P_i\, ,\\
\ddot\psi -\sum_{i=1}^9\dot\lambda_i^2&=&\frac 12 {{\rm e}^\psi} E\, ,
\end{array}
\label{motion1}
\end{equation}
where $\psi =2\phi-\sum_{i=1}^9\lambda_i$, is the shifted dilaton,
and  the energy, $E$, and pressure along the dimension $i$, $P_i$, depend
on the free energy of the string gas, $F$, as a function of $\beta$
 (the inverse temperature, $\beta =1/T$) and the scale factors $\lambda_i$,
\begin{equation}
\begin{array}{rcl}
E&=&-2\frac{\delta S_m}{\delta g_{00}}=F+\beta\frac{\partial F}{\partial\beta}\, ,\\
P_i&=&-\frac{\delta S_m}{\delta\lambda_i}=-\frac{\partial F}{\partial\lambda_i}\, .
\end{array}
\label{motion2}
\end{equation}
Finally, since we are interested in an adiabatic evolution, we should
impose that the adiabaticity condition $\frac{dS}{dt}=0$. It is worth 
pointing out that the equations of motion, eq.~(\ref{motion1}), and
 eq.~(\ref{motion2}) exhibit
{\it T duality} symmetry\cite{{Brandenberger:1988aj},{Veneziano:1991ek}}, 
$R\rightarrow 1/R$ \footnote{We have set $\alpha^\prime =1$.} ($\lambda_i\rightarrow -\lambda_i$, 
$\phi\rightarrow\phi -\sum_i\lambda_i$), being $R=1$ the dual radius, since
as we said before
 this gives rise to a non singular
cosmology.

The next step in the analysis is  the free energy. In analogy to
field theory, the free energy of a thermal gas of strings is calculated
by computing the one loop world-sheet path integrals for string propagation
on $R^{d-1}\times S$, where $S$ corresponds to the euclidean time compactified
in a circumference of radius $\beta$, which is equivalent to
require periodic bosons and anti-periodic fermions\cite{Atick:1988si}. 
Therefore, the calculation of $F$ is a one loop string calculation of the 
partition function of type IIA/B string theory compactified on
 $S_1/{\bf Z}_2$, where 
${\bf Z}_2$ is generated by $g$, the product of a  translation $\pi\beta$
 along the circle and $(-1)^{\cal F}$, with ${\cal F}$ the 
spacetime fermion operator. By means of the unfolding technique
 \cite{O'Brien:pn}, and considering
$9-d$ compact dimensions with radius, $r_i$, the free energy is simplified to 
\be
F(r_i ,\beta )=\frac{V_d}{2\pi}\int_{-1/2}^{1/2}d\tau_1\int_0^\infty d\tau_2\,\frac{1}{\tau_2^{(3+d)/2}} \prod_{i=d+1}^9\Lambda (r_i,\tau )\sum_{p=1}^\infty |M^2(\tau )| \,
{\rm e}^{-\frac{\beta ^2p^2}{4\pi\tau_2}}\, ,
\label{free}
\ee
where $\tau$ is the modular parameter of the torus, and the lattice contribution and the spin structure are given by 
\bea
\Lambda (r_i,\tau )&=&\sum_{m,n=-\infty}^\infty
q^{\frac 14 \left(\frac{m}{r_i}+nr_i \right)^2} 
\bar q^{\frac 14 \left(\frac{m}{r_i}-nr_i \right)^2}\, ,\\
|M^2(\tau )|&=&{\left (\frac{\theta_2^4 (\tau )}{\eta ^{12} (\tau )} \right )^2}=\sum_{N,\bar N =0}^\infty 
D(N)D(\bar N)q^Nq^{\bar N};
\eea
where $q={\rm e}^{2\pi i\tau}$, $D(N)$ is the degeneracy factor at
string level $N$, $m$/$n$ are the KK/winding mode along $r_i$,
 and $\theta_2$ and $\eta$ are modular functions
on the torus\cite{PolBook}.
Eq.~(\ref{free}) diverges for $\beta >\beta_H$, being 
 $\beta_H=2\sqrt2\pi$ the inverse of the Hagedorn temperature\cite{Hage}. In order to
manipulate eq.~(\ref{free}), we should carefully consider the regime we are 
working in, since when dealing with free string dynamics in the canonical ensemble,
fluctuations may diverge as the Hagedorn temperature is approached\cite{Mitchell:hr}, 
invalidating the thermodynamic limit. Therefore in this situation, corresponding to high
energy densities, the micro-canonical string ensemble is more physically appropriate.
On the other hand, as the energy fluctuations are small compared to the
average energy, lower energy densities, we can trustfully work with the canonical
 ensemble, which is easier to manipulate.

\section{Hagedorn regime}

In this regime we are dealing with high energy densities and therefore
to correctly study the thermodynamics we must make use of the micro-canonical
string ensemble. The main quantity to study is the energy density of
states $\Omega (E)$\cite{DJT}, which is governed by the analytic structure of the
canonical partition function, $Z(\beta )={\rm Tr}\,{\rm e}^{-\beta\, F}$, in the complex $\beta$ plane,
\be
\Omega (E)=\sum_\alpha\delta (E-E_\alpha )
=\int_{L-i\infty}^{L+i\infty}\frac{d\beta}{2\pi i}\, {\rm e}^{\beta E}\, Z(\beta ).
\ee
In this regime we consider all 9 small compact spatial dimensions, with the same
 radius $r_i=r$ in eq.~(\ref{free}).
In order to get an $r$ dependence we need to include, in addition to the 
 leading singularity at $\beta_H$,  the first next-to-leading
 singularities at $\beta_W=2\sqrt 2 \pi
(1-\frac{1}{2r^2})^{1/2}$ and $\beta_K=2\sqrt 2 \pi (1-\frac{r^2}{2})^{1/2}$.
The partition function, can then be parametrized as
\be
Z(\beta) = 
\frac{{\rm e}^{\Lambda(\beta,r)}}{\beta-\beta_H}\bigg(\frac{\eta_K}{\beta-\beta_K} 
\bigg)^{18}\bigg(\frac{\eta_W}{\beta-\beta_W} \bigg)^{18}\, ,
\ee
where $\eta_{K/W}=\beta_H-\beta_{K/W}$.
Given this, the calculation of the entropy and the thermodynamical quantities
is straightforward. For radius close to the dual radius the energy and pressure of the ideal gas
of strings take the form
\bea
\frac 1T &\sim& \frac{1}{T_H}+C_1 E^{17}
{\rm e}^{-\eta E}\, ,\nonumber\\
P &\sim& C_2E^{17}{\rm e}^{-\eta E}\, ,
\eea
where $C_1$ and $C_2$ are polynomial functions of the singularities, 
 $\eta \approx \sqrt 2\pi [2-(4-\frac{2}{r^2})^{1/2}]$ and $T_H=1/\beta_H$. 
As we see, temperature and pressure exhibit an exponential suppression
coming from the energy and therefore at high energy densities 
the pressure is negligible and  the temperature is nearly  constant 
 corresponding to the Hagedorn one. 
The string gas then, in this regime, presents an  equation of state 
corresponding to that of   pressureless dust and 
 then, the time evolution of the dimensions is such that they remain nearly 
constant close to the dual radius. In other words, in order that
the dimensions present interesting dynamics,
 such as expansion or contraction, the system must get out of thermal
equilibrium in accordance with B\&V model.

We can  impose as well the conservation of the total winding number $n_i$ and 
the discrete momenta $m_i$ along each compact spatial dimension by introducing
a chemical potential $\mu$ for each conserved charge $Q$\cite{DJT2}, such that 
$Z(\beta ,\mu )={\rm Tr}\,{\rm e}^{-\beta\, F +2\pi i\mu Q}$, with
 $\mu Q\rightarrow
\mu_i m_i+\nu_i n_i$. In contrast with the case without conservation, already
the leading singularity  depends on the compactification radius,  
$2\beta_H (\mu_i ,\nu_i, r_i ) = \sqrt{\beta_0^2 + \sum_i(\frac{\mu_i}{r_i}+\nu_i r_i)^2 }
+ \sqrt{\beta_0^2 + \sum_i(\frac{\mu_i}{r_i}-\nu_i r_i)^2 }$,
where $\beta_0 =2\pi\sqrt 2$. However, for high energies the pressureless
dust behavior is recovered.

\section{Almost radiation regime}

In the previous section, once the system gets out of thermal
equilibrium it is possible that interesting dynamics arise.
For example,  annihilation of winding and anti-winding states 
would take place
in some dimensions letting them free to expand, whereas the other 
dimensions would be  unable to expand due to the windings around them.
If we suppose such a mechanism took place, we
 could ask how the dynamics are once
the system gets again into thermal equilibrium. More explicitly,
if we start off with some large and some small dimensions,
how is their evolution in thermal equilibrium? Is it possible to
keep this hierarchy on sizes? 

At this point, since we are dealing with lower temperatures, we are
able to  use
 the canonical string ensemble, which deals with the canonical
partition function. The realization of  $d$ large and $9-d$ small
 dimensions
in the calculation, is given by the fact 
that only the small compact dimensions feel the
 contribution of winding and KK modes in the partition function.
 The free energy, eq.~(\ref{free}), can be split
in {\it a)} the contribution from the zero modes ($N=\bar N=m_i=n_i=0$) which 
we refer to as {\it radiation}
contribution, {\it b)} {\it matter} contribution coming from the non-zero 
modes ($N,\bar N,  m_i, n_i \neq 0$).

Concerning the radiation contribution it turns out that only the 
large dimensions do feel it. More explicitly, the large dimensions present
a radiation-like equation of state, $P^{\rm rad}_d=E^{\rm rad}/d$,
 whereas the pressure for the small dimensions is zero, $P^{\rm rad}_{9-d}=0$.
This translates into the fact that the large dimensions are able to
expand, independent of the initial conditions\footnote{Even for negative
initial expansion rate, the large dimensions initially contract, bounce and
finally expand.}, whereas the small dimensions stay around their 
initial value.

On the other hand,  the massive contribution affects both,  large and
small dimensions. For simplicity, we just considered the
lightest  KK and winding
modes along  the compact dimensions. The energy and pressures are 
given in terms of modified Bessel functions, which are functions of the
KK and winding modes,
\bea
E^{mat}&\sim&\frac{f(\beta )}{r^{(d+1)/2}} [ K_{(d+1)/2}\left(
\beta /r\right)+
\beta /r K^\prime_{(d+1)/2}\left(
\beta /r \right) \nonumber\\
&&  + r^{d+1} \{ K_{(d+1)/2}\left(\beta r\right) +
\beta r K^\prime_{(d+1)/2}\left(\beta r\right) \} ]\, , \\
P^{mat}_d&=& \frac{f(\beta )}{r^{(d+1)/2}}\left[ K_{(d+1)/2}
\left(\beta /r\right) +  r^{d+1} K_{(d+1)/2}\left(\beta r\right)\right ] \, ,
\label{prel}\\ 
P^{mat}_{9-d}&=&\frac{g(\beta )}{r^{(d+3)/2}}\left[ K_{(d-1)/2}\left(
\beta /r\right) -   r^{d+3} K_{(d-1)/2}\left(\beta r\right) \right] \, ,
\label{pres}
\eea
where $f(\beta )$ and $g(\beta )$ are functions of $\beta$ and the prime
denotes derivative with respect to $r$. As seen in eq.~(\ref{prel}) and 
eq.~(\ref{pres}), the winding modes  lead to a positive contribution
to the large dimensions and to a negative 
to the small ones, contrary to the KK modes which give positive contribution
to all dimensions.

\begin{figure}[t]
\epsfxsize=18pc 
\begin{center}
\epsfbox{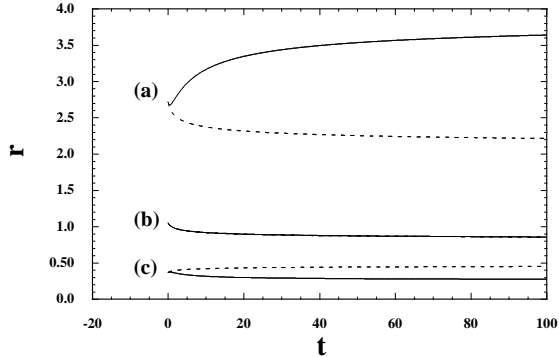}
\caption{Evolution of the small dimensions in the almost radiation regime.
 a) $\nu_0=1,\,\dot\nu_0=-0.1,\,$ b) $\nu_0=0.05,\,\dot\nu_0=-0.1,\,$ 
c) $\nu_0=-1.0,\,\dot\nu_0=0.1,\,
\psi_0=-16,\,\beta_0=12.$} 
\end{center} 
\label{fig}
\end{figure}
Whereas eq.~(\ref{prel}) always leads to an expansion of the large dimensions,
the effect of eq.~(\ref{pres}) depends on the initial size of the small 
dimensions. For initial radius larger than the dual radius, 
 (a) in fig.~\ref{fig}, the pressure,  
eq.~(\ref{pres}), is such that it  makes the small dimensions  expand.
In other words, the probability of having winding modes along
the compact dimensions is smaller as the radius is larger, letting the
 dimensions
 free to expand.
 On the other hand, for initial radius less than the
dual one, the probability of having winding modes  is larger, 
  leading to the contraction of the small dimensions, (b) in fig.~\ref{fig}.
 Finally, for radius close to the dual one\footnote{Thinking about B\&V model,
this is the case we are interested in, since the small dimensions are
supposed to be around the self dual radius.}, (c) in fig.~\ref{fig}, the effect of the winding modes
is compensated by that of the KK modes resulting in  zero 
pressure, and  giving rise to the stabilization of the small dimensions. Fig.~(\ref{fig}),
 shows the behavior for the different
choice of initial conditions with the contribution of  pure radiation
(dotted lines) and the contribution of pure radiation and matter (solid lines).

We therefore conclude that in the almost radiation regime, as long as the
small dimensions start close to the self dual radius, they remain small
whereas the large dimensions expand. It is good to point out that
these results are independent of the number of large/small dimensions. Finally,
as in the Hagedorn case, imposing the conservation of KK and winding modes
does not change the picture quantitatively.

\section*{Acknowledgments}

I would like to thank  S. Tsujikawa, B. Bassett and specially M. Serone for their
collaboration in the work this talk is based on.
 I am  also happy to thank M. Blau for useful discussions.


\begin{thebibliography}{99}

\bibitem{Horowitz:2002mw}
G.~T.~Horowitz and J.~Polchinski,
Phys.\ Rev.\ D {\bf 66} (2002) 103512.

\bibitem{Craps:2002ii}
B.~Craps, D.~Kutasov and G.~Rajesh,
JHEP {\bf 0206} (2002) 053.

\bibitem{Liu:2002ft}
H.~Liu, G.~Moore and N.~Seiberg,
JHEP {\bf 0206} (2002) 045.\\
H.~Liu, G.~Moore and N.~Seiberg,
JHEP {\bf 0210} (2002) 031.

\bibitem{Quevedo:2002xw}
F.~Quevedo,
Class.\ Quant.\ Grav.\  {\bf 19} (2002) 5721.

\bibitem{Lidsey:1999mc}
J.~E.~Lidsey, D.~Wands and E.~J.~Copeland,
Phys.\ Rept.\  {\bf 337}, 343 (2000).


\bibitem{Brandenberger:1988aj}
R.~H.~Brandenberger and C.~Vafa,
Nucl.\ Phys.\ B {\bf 316} (1989) 391.


\bibitem{Gasperini:2002bn}
M.~Gasperini and G.~Veneziano,
Phys.\ Rept.\  {\bf 373} (2003) 1.

\bibitem{Khoury:2001bz}
J.~Khoury, B.~A.~Ovrut, N.~Seiberg, P.~J.~Steinhardt and N.~Turok,
Phys.\ Rev.\ D {\bf 65} (2002) 086007.

\bibitem{Binetruy:1999ut}
P.~Binetruy, C.~Deffayet and D.~Langlois,
Nucl.\ Phys.\ B {\bf 565} (2000) 269.

\bibitem{Bassett:2003ck}
B.~A.~Bassett, M.~Borunda, M.~Serone and S.~Tsujikawa,
Phys.\ Rev.\ D {\bf 67} (2003) 123506.

\bibitem{Watson:2003gf}
S.~Watson and R.~Brandenberger,
arXiv:hep-th/0307044.


\bibitem{Gasperini:1992em}
M.~Gasperini and G.~Veneziano,
Astropart.\ Phys.\  {\bf 1} (1993) 317.

\bibitem{Veneziano:1991ek}
G.~Veneziano,
Phys.\ Lett.\ B {\bf 265}, 287 (1991); \\
K.~A.~Meissner and G.~Veneziano,
Phys.\ Lett.\ B {\bf 267}, 33 (1991).

\bibitem{TV}
A.~A.~Tseytlin and C.~Vafa,
Nucl.\ Phys.\ B {\bf 372} (1992) 443.


\bibitem{Atick:1988si}
J.~J.~Atick and E.~Witten,
Nucl.\ Phys.\ B {\bf 310} (1988) 291.

\bibitem{O'Brien:pn}
K.~H.~O'Brien and C.~I.~Tan,
Phys.\ Rev.\ D {\bf 36} (1987) 1184;\\
E.~Alvarez and M.~A.~Osorio,
Phys.\ Rev.\ D {\bf 36}, 1175 (1987). 

\bibitem{PolBook}
J.  Polchinski, {\it String Theory}, Volume 1, Cambridge, page. 214-216.

\bibitem{Hage}
R. Hagedorn, Suppl. Nuovo Cimento {\bf 3} (1965) 147.

\bibitem{Mitchell:hr}
D.~Mitchell and N.~Turok,
Phys.\ Rev.\ Lett.\  {\bf 58} (1987) 1577.

\bibitem{DJT}
N.~Deo, S.~Jain and C.~I.~Tan,
Phys.\ Lett.\ B {\bf 220} (1989) 125.

\bibitem{DJT2}
N.~Deo, S.~Jain and C.~I.~Tan,
Phys.\ Rev.\ D {\bf 40} (1989) 2626.

\end{thebibliography}
\end{document}